\documentclass{emulateapj}
\usepackage{natbib,times}
\usepackage{lscape}
\slugcomment{Received 2011 January 2; accepted 2011 April 7; published 2011 May 25 on ApJ 734, 68.}

\shorttitle{Galaxy clusters at high redshift}
\shortauthors{Wen \& Han}
\begin{document}

\title{Galaxy clusters at high redshift and evolution of brightest cluster galaxies}

\author{Z. L. Wen, \and J. L. Han}
\affil{National Astronomical Observatories, 
                Chinese Academy of Sciences, 
                20A Datun Road, Chaoyang District, Beijing 100012, China; 
                zhonglue@nao.cas.cn.} 


\begin{abstract}
Identification of high redshift clusters is important for studies of
cosmology and cluster evolution. Using photometric redshifts of
galaxies, we identify 631 clusters from the Canada-France-Hawaii
Telescope (CFHT) Wide field, 202 clusters from the CHFT Deep field,
187 clusters from the Cosmic Evolution Survey (COSMOS) and 737
clusters from the Spitzer Wide-area InfraRed Extragalactic survey
(SWIRE) field. The redshifts of these clusters are in the range of
$0.1\lesssim z\lesssim1.6$. Merging these cluster samples gives 1644
clusters in the four survey fields, of which 1088 are newly identified
and more than half are from the large SWIRE field. Among 228 
clusters of $z\ge1$, 191 clusters are newly identified,
and most of them from the SWIRE field.
With this large sample of high redshift clusters, we study the color
evolution of the brightest cluster galaxies (BCGs). The colors $r'-z'$
and $r^+-m_{3.6\mu m}$ of the BCGs are consistent with a stellar
population synthesis model in which the BCGs are formed at redshift
$z_f\ge2$ and evolved passively. The colors $g'-z'$ and $B-m_{3.6\mu
  m}$ of the BCGs at redshifts $z>0.8$ are systematically bluer than
the passive evolution model for galaxy formed at $z_f\sim2$,
indicating star formation in high redshift BCGs.
\end{abstract}


\section{Introduction}

Galaxy clusters are important objects to study the large scale
structure \citep{bah88,phg92} and constraint the cosmological
parameters, e.g., $\Omega_m$, the mass density parameter of the
universe, and $\sigma_8$, the amplitude of mass fluctuations at a
scale of 8 $h^{-1}$ Mpc \citep{bfc97,rb02,whl10}.
Clusters are also important laboratories to investigate the properties
of dark matter, hot gas, galaxies and active galactic nucleus in
dense environment \citep[e.g.,][]{cs03,dre80,cdn07}.
However, most of these investigations are based on clusters in the local
universe. 

High redshift clusters can provide information on cosmological
structure in the vast universe and the evolution of cluster properties
with cosmic time. It has been shown that cluster galaxies at high
redshift have more star formation than those at low redshift
\citep{bo78,bo84}. The evolution of cluster mass functions gives
constraints on not only the cosmological parameters but also dark
energy equation of state parameter $\omega_0$
\citep[e.g.,][]{vkb+09}. The abundance of (even a few) high
redshift massive clusters constrains the non-Gaussianity of primordial
perturbation \citep{mvj00,hjv10}. High redshift clusters are needed to
study the formation and evolution of brightest cluster galaxies
\citep{moz+06,wad+08,ses+08}.

\subsection{Previous detection of high redshift clusters}

Tens of thousands of galaxy clusters have been identified from various
surveys in the last decades \citep[e.g.,][]{aco89,whl09,hmk+10}. They
mostly have redshifts $z<0.5$. High redshift clusters were recently
identified from multi-color optical or infrared deep surveys. Only a
few hundreds of clusters have redshifts $z>1$
\citep[e.g.,][]{gy05,ghi+08}.

Using cluster red sequence method, \citet{gy05} detected 429 candidate
clusters or groups with redshifts of $0.2<z<1.4$ from the Red-Sequence
Cluster Survey (RCS), of which 67 have $0.9<z<1.4$. Using the Cut and
Enhance method, \citet{ghi+08} identified 16 cluster candidates at
redshifts $0.9<z<1.7$ from the AKARI deep survey. \citet{ebg+08} used
a wavelet algorithm based on photometric redshift to the Spitzer
Infrared Array Camera (IRAC) Shallow Survey and NOAO Deep Wide-Field
Survey data and identified 335 cluster and group candidates, of
which 106 have redshifts $z\ge 1$. 

\citet{obc+07} applied a matched-filter cluster detection algorithm to
the Canada-France-Hawaii Telescope Legacy Survey (CFHTLS). They
identified 162 cluster candidates over an area of 3.112 deg$^2$, of
which 29 have redshifts $z\ge 1$. \citet{gbm09} found 114 cluster
candidates by using a matched filter detection method to the CFHTLS
$r'$ band data of the Deep fields, and 247 cluster candidates from the
$z'$ band data. Merging the samples in both bands gives 233 clusters,
of which 93 have redshifts $z\ge 1$. \citet{twc09} found 5804 cluster
candidates from the CFHT Wide field of 161 deg$^2$ using the red
sequence method, of which 13 have redshifts $z\ge 1$.  By mapping
density of galaxy distribution in photometric redshift space,
\citet{adb+10} detected 1200 candidate clusters with masses greater
than $1.0\times10^{13}~M_{\odot}$ from the CFHTLS Deep and Wide
fields, of which 302 have redshifts $z\ge1$. \citet{mvh+10} applied a
3-dimensional matched-filter cluster detection algorithm to the CFHTLS
data and identified 673 cluster candidates in the redshift range of
$0.2\le z\le1.0$.

\citet{kli+09} used the friends-of-friends and Voronoi tessellation
methods and identified 102 groups of $z\le 1$ with more than five
member galaxies from $\sim$10,000 redshifts from the Cosmic Evolution
Survey (COSMOS) field. \citet{vcb+06} identified 13 clusters in the
redshift range of $0.61\le z\le1.39$ from the UKIRT Infrared Deep Sky
Survey (UKIDSS) Early Data Release. \citet{zrw+07} identified 12
cluster candidates in the redshift range of $1.23\le z\le1.55$ from
the Heidelberg InfraRed/Optical Cluster Survey (HIROCS) survey in the
COSMOS field. \citet{ccd+10} found three candidate clusters of
galaxies at redshifts most likely between 1.7 and 2.0 in the COSMOS
field. \citet{wph+10} used the friend-of-friend method to the Great
Observatories Origins Deep Survey (GOODS) data and identified 206
group at redshift between 0.4 and 1.0. From the first 36 XMM-Newton
pointings on the COSMOS field, \citet{fgh+07} identified 72 clusters,
of which 8 have redshifts $z\ge 1$. From the XMM-LSS survey,
\citet{pcp+07} identified 73 clusters. From the Subaru-XMM Deep Field,
\citet{fwt+10} identified 57 cluster candidates, of which 13 have
redshifts $z\ge 1$.

Besides the sample finding, some authors have spectroscopically found
or confirmed individual high redshift clusters. \citet{kcz+09}
confirmed a galaxy cluster of $z=1.6$ from the Galaxy Mass Assembly
ultra-deep Spectroscopic Survey (GMASS). \citet{gye+08} confirmed a
compact supercluster structure comprising three clusters at
$z=0.9$. \citet{pmw+10} discovered a galaxy cluster at $z=1.62$ from
the Spitzer Wide-Area Infrared Extragalactic survey (SWIRE) XMM-LSS
field. \citet{app+08} confirmed a cluster at $z=1.016$ using a
modified red sequence method, follow up spectroscopy and X-ray
imaging. From the Spitzer/IRAC Shallow Survey of the Bootes field,
\citet{seb+05} confirmed a galaxy cluster at $z=1.41$. From the
Spitzer Adaptation of the Red-sequence Cluster Survey (SpARCS),
\citet{mwy+09}, \citet{wmy+09} and \citet{dwm+10} confirmed two
clusters at $z\sim1.2$, one cluster at $z=1.34$ and three cluster at
$z=0.87$, 1.16, 1.21, respectively. Using a 3-dimensional technique,
\citet{cst+07} detected a forming galaxy cluster at redshift 1.6 in
the GOODS field. A few high redshift massive clusters were found by
X-ray observations and Sunyaev--Zeldovich (SZ) effect. From the
XMM-Newton observations, a massive X-ray cluster at $z=1.39$ was
discovered by \citet{mrl+05} and a massive X-ray cluster at $z=1.45$
by \citet{srs+06}, and a cluster at $z=1.22$ was identified by
\citet{bvw+06} and a cluster at $z=0.95$ by \citet{fbl+08}. Using the
SZ effect, \citet{bra+10} found a massive cluster at $z=1.07$ from the
South Pole Telescope (SPT) data.

\subsection{Evolution of brightest cluster galaxies}

The brightest cluster galaxy (BCG) is an elliptical galaxy located at
the potential center of a galaxy cluster. BCGs are the most luminous
galaxies in the universe and {\it in general} have no prominent
ongoing star formation. They are red in the rest-frame color. Their
surface brightness profiles are different from those of ordinary
elliptical galaxies (non-BCGs), and they do not follow the basic
scaling relations of normal ellipticals
\citep[e.g.,][]{lfr+07,lxm+08}. Because of the dominant role inside
clusters and their unusual properties, the formation and evolution of
BCGs are very intriguing.

The properties and evolution of BCGs were studied using identified
high redshift clusters. \citet{ber09} noticed that BCGs at lower
redshift have larger sizes and smaller velocity dispersions,
indicating that early-type BCGs grow from many dry minor mergers
\citep{moz+06} rather than a few major mergers \citep{rfv07}. By
comparing the structural parameters of BCGs of nearby clusters
($0.04<z<0.07$) and those at intermediate redshift ($0.3<z<0.6$),
\citet{aav+11} confirmed the size decrease of BCGs with redshift. The
properties of BCGs are related to the host clusters. More luminous,
larger and more centrally located BCGs are located in more massive and
rich galaxy clusters.

On the $K$-band Hubble diagram, BCGs do not exhibit any luminosity
evolution with redshift \citep{cm98,abk98,wad+08} and significant
change in the stellar mass \citep{wad+08}. The colors of BCGs are in
good agreement with the evolved old stellar population formed at
$z>2$. This means that the stellar population in BCGs has been in
place since at lease $z=2$ \citep{ses+08} and the average stellar mass
of BCGs remains constant since $z\sim1.5$. \citet{scs+10} therefore
concluded that dry merger seems to have little effect on evolutions
of BCG stellar mass over the last 9 -- 10 Gyr.

However, \citet{lmd+09} found that dry merger plays an important role
in the stellar mass assembly of BCGs. \citet{db07} concluded from
hierarchical simulations that the half mass of a typical BCG is assembled
by dry mergers after $z\sim0.5$.  The stellar populations of BCGs are
formed very early, 50\% at redshift $z\sim5$ and 80\% at $z\sim3$.

Observationally, a very small number of BCGs, e.g., those in Abell
1835, Zw3146 and MACS J0913.7+4056, exceptionally show the features of
star formation \citep{mrb+06,erf+06,bhb+08,obp+08,dbw+10,hmd10} or
post-starburst features \citep{lwhm11}. The star formation is the
dominant power for the infrared and H$\alpha$ emission detected from
BCGs \citep{obp+08}.

As shown above, many controversies exist on the formation of BCGs and
the role of dry merger \citep{lmd+09,scs+10}. It is not clear when
the BCGs are formed, what process dominates the evolution of
BCGs. Finding more high redshift clusters and their BCGs is crucial to
study these questions.

In this paper, we identify high redshift clusters from the CFHT Wide
field, the CHFT Deep field, the COSMOS field and the Spitzer SWIRE
field. In Section 2, we first apply our cluster detection algorithm to
the photometric redshift data of the four fields, and identify
high-redshift clusters. We then discuss the false detection rate, the
accuracy of determined cluster redshift, cluster richness and X-ray
luminosity of some clusters. The color evolution of BCGs is studied in
Section 3. Conclusions of this paper are presented in Section 4.

Throughout this paper, we assume a $\Lambda$CDM cosmology by taking
H$_0=$100 $h$ ${\rm km~s}^{-1}$ ${\rm Mpc}^{-1}$, with $h=0.72$,
$\Omega_m=0.3$ and $\Omega_{\Lambda}=0.7$.

\begin{deluxetable*}{lcrccclcccc}
\tablecaption{Details of the four deep surveys used for identification
  of clusters.
\label{deepfields}
}
\tablecolumns{11}
\tablehead{
\colhead{name}&\colhead{field}     &\colhead{mag.}& \colhead{No. of} & \colhead{$\sigma_{\Delta z}/(1+z)$} & 
\colhead{photo-z} &\colhead{$M_{\rm lim}$} & \colhead{No. of} & \colhead{No. of}& \colhead{No. of high-$z$} & \colhead{No. 
of high-$z$} \\
               &\colhead{(deg$^2$)}&\colhead{limit} & \colhead{bands} &  & ref. & \colhead{this paper} & 
                \colhead{clusters} & \colhead{new clusters} & \colhead{clusters} & \colhead{new clusters}
}
\startdata 
CFHT Wide & 35.0  &  $i'<22.5$&5 & 0.038 & 1   &$M_{i'}\le -21.5$&631&285&24 &19\\
CFHT Deep & 3.2   &  $i'<24$  &5 & 0.029 & 1,2 &$M_{i'}\le -21.5$&202& 55&53 &29\\
COSMOS    & 2.0   &  $i^+<25$ &30 & 0.012& 3   &$M_V\le -20.5$   &187&127&43 &38\\
SWIRE     &33.0   &  $r<24$   &5--12&0.035&4   &$M_B\le -20.5$   &737&674&116&106
\enddata
\tablecomments{References for photo-$z$: (1). \citet{cik+09}; 
(2). \citet{iam+06}; (3). \citet{ics+09}; (4). \citet{rbo+08}.
Here, high-$z$ means $z\ge1$.
}
\end{deluxetable*}

\section{High redshift clusters detected from deep fields}

To identify high redshift clusters, one has to discriminate the
cluster member galaxies from field galaxies in deep fields. Note that
the projection effect is very serious in photometric data. The ``color
cut'' \citep{ghi+08} or ``red sequence'' \citep{gy05} methods have
been previously used for de-projection. Galaxy clusters can be
detected in a certain redshift range when proper filters covering the
4000\AA~ break feature are used for photometry. When the break feature
shifts out from the chosen filters, photometric data of other bands
must be used for color cuts. Cluster richness determined by different
color cuts may not match well \citep[e.g.][]{hmk+10}.  The enhanced
star formation in galaxies of high redshift clusters (i.e. the
Butcher-Oemler effect) also makes the cluster detection more difficult
by the color cuts.

The photometric redshifts of galaxies (photo-$z$) are determined 
by comparison of the multi-band photometric data with the spectral
energy distribution of galaxies including the 4000\AA~ break feature
\citep[e.g.][]{cbc+03}. The photometric redshifts have 
successfully been used for the de-projection
\citep[e.g.,][]{ebg+08,whl09,mvh+10,adb+10,gh11}.  When photometric
redshifts of galaxies are used for identification of clusters, one
does not have to consider the color of galaxies and the redshift
range. Here, we follow and modify the method described in
\citet{whl09} to identify high redshift clusters from the CFHT Wide
field, the CFHT Deep field, the COSMOS field and the Sptizer SWIRE
field. 

\subsection{Photometric redshift data for deep fields}

The photometric redshifts of galaxies have been estimated for many
multi-color deep surveys, e.g., the GOODS Southern Field
\citep{mib+04}, the RCS \citep{hyl+05}, the Hubble Ultra Deep Field
\citep{cbs+06}, the Spitzer IRAC Shallow Survey \citep{bba+06}, CFHT
\citep{iam+06}, SWIRE \citep{rbo+08}, COSMOS \citep{ics+09}, the ESO
Distant Cluster Survey field \citep{prd+09}, the AKARI survey
\citep{nsp+09}, the Multi-wavelength Survey by Yale-Chile
\citep{cvu+10}. These surveys provide accurate photometries of 4 to 32
bands from UV to mid-infrared, so that the estimated photometric
redshifts have an uncertainty of 0.01 -- 0.1. For cluster detection,
we restrict the photo-$z$ data with a small uncertainty of
$\sigma_{\Delta z}\le 0.04(1+z)$. In this paper, we work on the deep
fields larger than 1 deg$^2$. We find that photo-$z$ data for the
CFHT Wide field, the CFHT Deep field, the COSMOS field and the Sptizer
SWIRE field are available for cluster identification at high redshifts
(see details in Table~\ref{deepfields}).

The CFHT survey is carried out with the 3.6 m Canada-France Hawaii
Telescope in the $u'$, $g'$, $r'$, $i'$ and $z'$ bands for three
imaging sub-surveys: the CFHT Deep, the CFHT Wide and the CFHT Very
Wide. The photo-$z$ data of the Deep and Wide fields have been
published and used in previous work (see references in Sect.~1.1). The
CFHT Wide survey covers 35 deg$^2$ in three deep fields (W1, W3,
W4). The CFHT Deep survey covers 3.2 deg$^2$ in four deep fields (D1,
D2, D3, D4).

The COSMOS is made for a field of 2 deg$^2$ by many observations from
X-ray to radio (XMM, Galaxy Evolution Explorer, the Hubble Space
Telescope, CFHT, United Kindon Infrared Telescope, Subaru, Spitzer and
VLA). 

The Spitzer SWIRE is to provide photometry \citep{lsr+03} in the IRAC
bands of 3.6 $\mu$m, 4.5 $\mu$m, 5.8 $\mu$m and 8.0 $\mu$m, the MIPS
bands of 24 $\mu$m, 70 $\mu$m and 160 $\mu$m for six regions:
ELAIS-N1, ELAIS-N2, Lockman Hole, XMM, ELAIS-S1 and Chandra Deep Field
South (CDFS). The photometries in optical bands $u$, $g$, $r$, $i$ and
$z$ (only here is the band, not redshift) are available from the SWIRE
photometry programme and from \citet{mwi+01} and \citet{pcp+07}.

Detailed parameters for the four fields are given in
Table~\ref{deepfields}, including the field size, magnitude limit,
number of bands for photometry, and the accuracy of photo-$z$.

\subsection{Finding high redshift clusters}

\begin{deluxetable*}{rrrcrccrccl}
\tablecolumns{11}
\tablewidth{0pc}
\tablecaption{Clusters identified from the four deep fields.
\label{cat}}
\tablehead{ 
\colhead{Name}&\colhead{R.A.$_{\rm BCG}$}&\colhead{Decl.$_{\rm BCG}$} & \colhead{$z_p$}  & \colhead{$z_{s,\rm BCG}$}  &
\colhead{mag$_{\rm BCG}$}  & \colhead{$D$} & \colhead{N$_{\rm gal}$} & \colhead{$R$}& 
\colhead{$L_{\rm tot}$}   & \colhead{Prev.} \\
  &\colhead{(deg)}& \colhead{(deg)} & & & & & & &\colhead{($10^{10}L_{\odot}$)} &\colhead{catalog} \\
\colhead{(1)} & \colhead{(2)} & \colhead{(3)} & \colhead{(4)} & \colhead{(5)} & 
\colhead{(6)} & \colhead{(7)} & \colhead{(8)} & \colhead{(9)} & \colhead{(10)} &
\colhead{(11)} 
}
\startdata 
CFHT-W J021408.5$-$053410& 33.53388&$ -5.59269$& 0.4128&$-1.0000$& 18.70& 5.77& 20& 16.00& 106.41& 1,2 \\   
CFHT-W J021527.6$-$053338& 33.86837&$ -5.55166$& 0.2848&$-1.0000$& 17.70& 9.84& 25& 23.57& 120.67& 1,2 \\   
CFHT-W J021447.5$-$053309& 33.69700&$ -5.55293$& 0.8629&$-1.0000$& 20.83& 5.25& 25& 21.00& 186.48&     \\   
CFHT-W J021242.9$-$053219& 33.17939&$ -5.54365$& 1.1462&$-1.0000$& 22.23& 4.55& 14&  7.43&  95.94&     \\[1mm]
CFHT-D J022435.6$-$045505& 36.13370&$ -4.91589$& 0.9846&$-1.0000$& 21.26& 4.30& 29& 19.76& 199.19&     \\ 
CFHT-D J022723.5$-$045424& 36.82425&$ -4.91559$& 0.3135&$-1.0000$& 18.95& 4.28& 16& 14.31&  73.06& 1,2,3\\
CFHT-D J022624.6$-$045652& 36.59565&$ -4.95964$& 0.9416&$-1.0000$& 21.57& 5.14& 24& 19.71& 106.97& 4   \\
CFHT-D J022425.3$-$045229& 36.09802&$ -4.87084$& 0.9458&$-1.0000$& 21.13& 4.31& 32& 21.18& 190.84& 2,3,5\\[1mm] 
COSMOS J100313.1$+$013611&150.81335&   1.60198 & 0.5119&$-1.0000$& 20.25& 9.33& 15& 13.84&  49.84&     \\
COSMOS J100117.0$+$013618&150.34543&   1.62183 & 0.2283&$-1.0000$& 18.74& 7.25& 13& 11.37&  33.54&     \\
COSMOS J100112.4$+$013401&150.28815&   1.55581 & 0.3610&  0.3638 & 19.08& 6.47& 36& 32.15& 105.62& 6   \\
COSMOS J100157.6$+$020343&150.49001&   2.06934 & 0.4385&  0.4409 & 18.93& 7.60& 14& 12.29&  44.30& 3,6 \\[1mm] 
 SWIRE J003847.1$-$433227&  9.67174&$-43.54514$& 0.3609&$-1.0000$& 18.88& 4.46& 14&  8.00&  28.39&     \\
 SWIRE J003847.2$-$434822&  9.68008&$-43.79479$& 0.4804&$-1.0000$& 19.17& 7.57& 26& 17.11& 142.17&     \\
 SWIRE J003442.1$-$430746&  8.67131&$-43.14276$& 0.8072&$-1.0000$& 21.87& 8.44& 22& 16.00& 102.18&     \\
 SWIRE J022717.1$-$044557& 36.81958&$ -4.74909$& 1.3001&$-1.0000$& 24.04& 4.43& 16& 11.60&  69.21& 7          
\enddata
\tablecomments{ Same clusters found in different fields are repeatedly listed here for 
completeness with the same BCG coordinates. 
Column (1): Cluster name given by field name and J2000 coordinates of cluster center; 
Column (2): R.A. (J2000) of BCG; 
Column (3): Decl. (J2000) of BCG; 
Column (4): cluster redshift estimated from the median photo-$z$ of member galaxies;
Column (5): spectroscopic redshift of the BCG if available. `$-1.0000$' stands for not available;
Column (6): BCG magnitude at $i'$-band (7629 \AA) for the CFHT-W and CFHT-D fields,
$i^+$-band (7629 \AA) for the COSMOS field, and $r$-band (6230 \AA) for the SWIRE field;
Column (7): overdensity level; 
Column (8): number of member galaxy candidates within a radius of 1 Mpc and the redshift gap of $z\pm1.44\,\sigma_{\Delta z}$. See $\sigma_{\Delta z}$
 in Table~\ref{deepfields};
Column (9): cluster richness;
Column (10): total luminosity of cluster member galaxies, after the local background is subtracted. It is in the $i'$-band for the CFHT-W and CFHT-D fields, 
$V$-band for the COSMOS field, $B$-band for the SWIRE field;
Column (11): previous catalog containing the cluster: (1). \citet{twc09}; (2). \citet{adb+10}; (3). \citet{obc+07}; 
(4). \citet{gbm09}; (5). \citet{mvh+10}; (6). \citet{fgh+07} for X-ray; (7). \citet{pcp+07} for X-ray. \\ 
{\it This table is available in its entirety in a machine-readable form in the online journal. 
A portion is shown here for guidance regarding the form and content.}
}
\end{deluxetable*}

We identify galaxy clusters using the photo-$z$ data for
discrimination of {\it luminous} member galaxies from the four fields
with following steps \citep{whl09}:

1. Assuming that each galaxy at a given photometric redshift, $z$, is
the central galaxy of a cluster candidate, we count the number of
luminous member galaxies of $M \le M_{\rm lim}$, $N(0.5)$, within a
radius of 0.5 Mpc and a photo-$z$ gap of $z\pm1.44\,\sigma_{\Delta
  z}$. See the $M_{\rm lim}$ and $\sigma_{\Delta z}$ values in
Table~\ref{deepfields}. Here, the factor of 1.44 for the photo-$z$ gap
is chosen so that more than 85\% of member galaxies of a cluster can
be included assuming that the uncertainties of photometric redshifts
of member galaxies follow a Gaussian distribution. A larger $z$ gap
can help to include more member galaxies, but more background galaxies
are involved. We set the absolute magnitude limit, not only to make
our cluster detection consistent even up to high redshifts, but also
to exclude less luminous galaxies as member galaxies which have a
larger photo-$z$ uncertainty. The absolute magnitudes of galaxies at
the $i'$-band for the CFHT Wide and the CFHT Deep data, the $V$-band
for the COSMOS data and the $B$-band for the SWIRE data are provided
together with photo-$z$ data of galaxies in their catalogs
\citep{cik+09,ics+09,rbo+08}. The upper limits of absolute magnitude
for the four fields are also listed in Table~\ref{deepfields}. The
absolute magnitudes of galaxies are not available for the CFHT Deep
field in \citet{iam+06}, and we estimate the absolute magnitudes in
the $i'$-band using the K-correction of \citet{fsi95} according to the
types of galaxies given by \citet{iam+06}.

2. We apply the friend-of-friend algorithm to the luminous galaxies
using a linking length of 1 Mpc in the transverse direction and a
photo-$z$ gap of $z\pm3\,\sigma_{\Delta z}$. We then find the linked
galaxy with the maximum $N(0.5)$ as the temporary center of a cluster
candidate. If two or more galaxies show the same maximum number count,
the brightest one is taken as the temporary central cluster
galaxy. The linking length and photo-$z$ gap are large enough to link
all member galaxies in the transverse distance and redshift
space. This step can ensure to avoid multiple detection of a cluster.

\begin{figure}[bt]
\epsscale{1.1}
\plotone{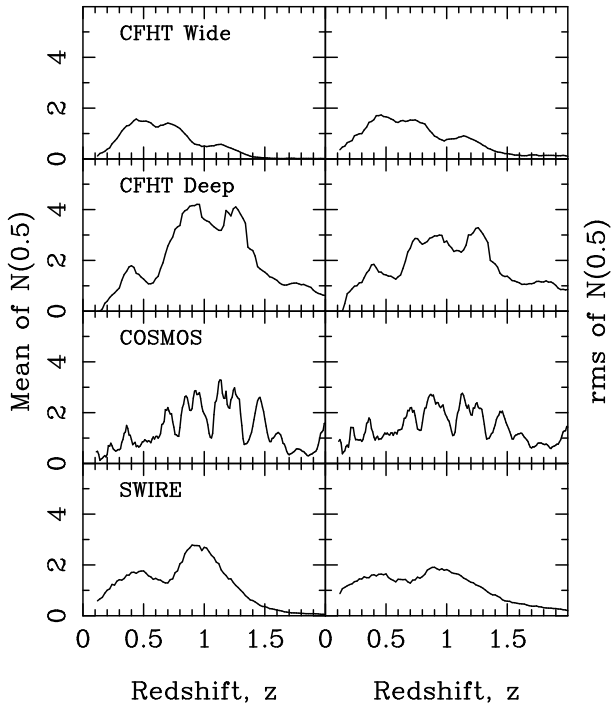}
\caption{Mean and rms of background galaxies of $M_r\le M_{\rm lim}$ 
within a radius of 0.5 Mpc and a redshift gap between $z\pm1.44\,\sigma_{\Delta z}$.
\label{bgc}}
\end{figure}

3. For each galaxy with maximum $N(0.5)$ at $z$, all galaxies within a
radius of 1 Mpc from the temporary central galaxy and the photo-$z$ gap
between $z\pm1.44\,\sigma_{\Delta z}$ are assumed to be the member
galaxies.  The cluster redshift $z_p$ is then defined to be the median
value of the photometric redshifts of the recognized ``members''. The
absolute magnitudes of member galaxies are re-calculated with this
cluster redshift.

4. The galaxy overdensity of a cluster, $D=(N(0.5)-\langle
N(0.5)\rangle)/\sigma_{N(0.5)}$, is estimated from the member
number and the background galaxy density of the same magnitude
limit. Here, $\langle N(0.5)\rangle$ is the mean galaxy count of the
same magnitude limit within 0.5 Mpc for each field for the given
redshift gap, $\sigma_{N(0.5)}$ is the root mean square (rms) of the
number count (see Figure~\ref{bgc}). In practices, 300 random
positions (R.A., Decl.) are selected in the real data, and the number
of luminous galaxies of $M\le M_{\rm lim}$ is counted within a radius
of 0.5 Mpc and a redshift gap between $z\pm1.44\,\sigma_{\Delta
  z}$. The mean and the rms are statistically obtained from these 300
values.  A larger value of $D$ means a higher likelihood of a true
cluster (see Table \ref{cat}). We set the threshold, $D\ge4$, for
cluster identification. Meanwhile, we also require $N(0.5)\ge8$ to
avoid the detection of poor clusters or false detection in case that
the mean number and rms of background are very small. Now, if all
conditions are satisfied, a cluster is identified.

5. The real center of a detected cluster is then defined as the
average position of luminous member galaxy candidates of $M\le M_{\rm
  lim}$ within radius of 1 Mpc (not 0.5 Mpc) from the temporary center
and the redshift gap of $z\pm1.44\,\sigma_{\Delta z}$. This position
is given in the cluster name. The cluster richness, $R$, is defined as
the {\it real} number of cluster galaxies, which is obtained from the
number of member galaxy candidates, $N_{\rm gal}$, within a radius of 1 Mpc from
the real center of a cluster and the redshift gap subtracted by the
local contaminations $\langle N_{\rm cb}\rangle$ of foreground and
background galaxies, so that $R=N_{\rm gal}-\langle N_{\rm
  cb}\rangle$. The contamination, $\langle N_{\rm cb}\rangle$, is
estimated from the local distribution of the galaxies in this redshift
gap \citep[see details in][]{whl09}.

6. The total luminosity of each cluster, $L_{\rm tot}$, is calculated
as the total luminosity of member galaxies within the region after the
similar subtraction of the average background contribution.

\begin{figure}
\resizebox{42mm}{!}{\includegraphics{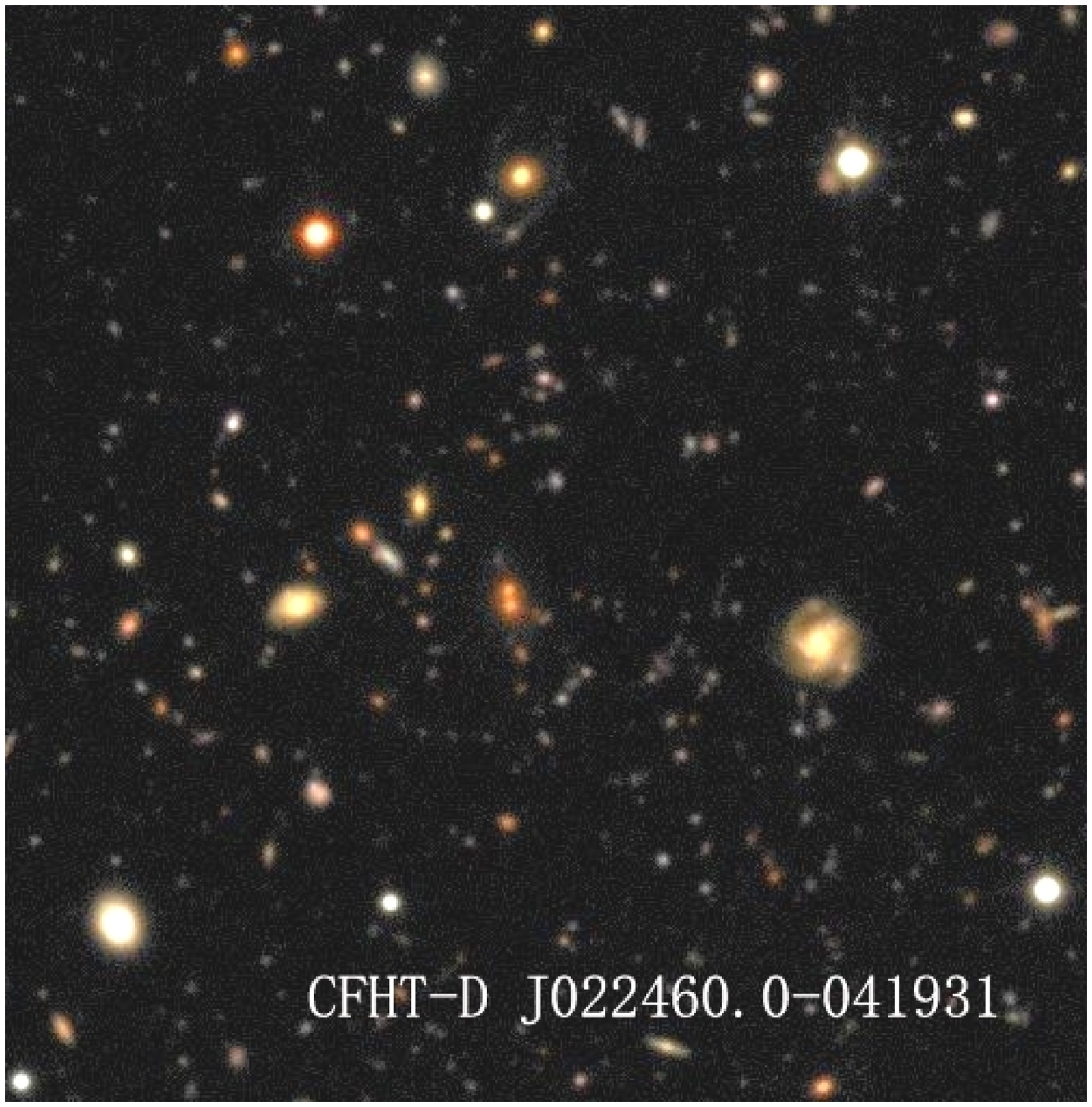}}
\resizebox{42mm}{!}{\includegraphics{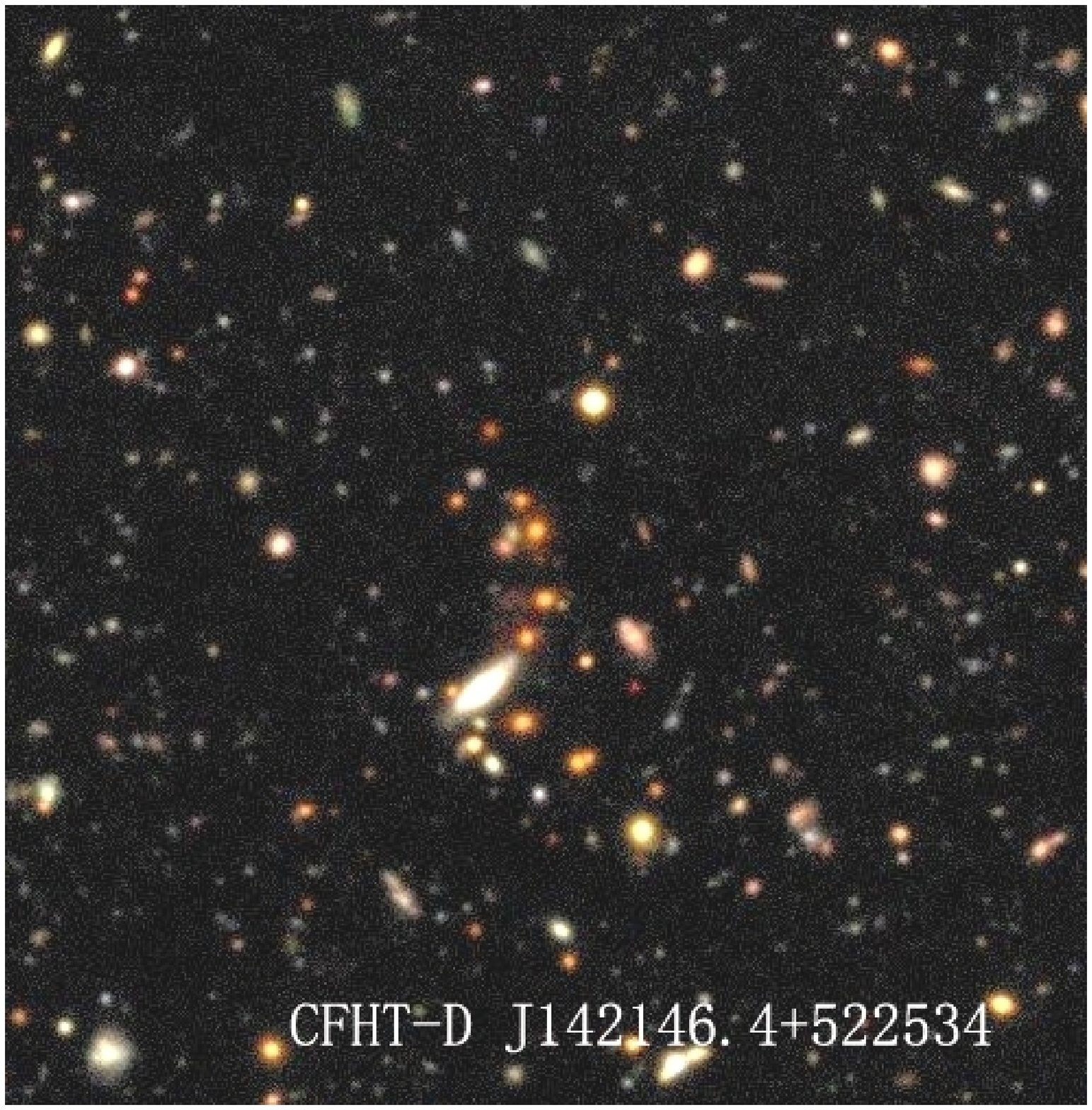}}\\
\resizebox{42mm}{!}{\includegraphics{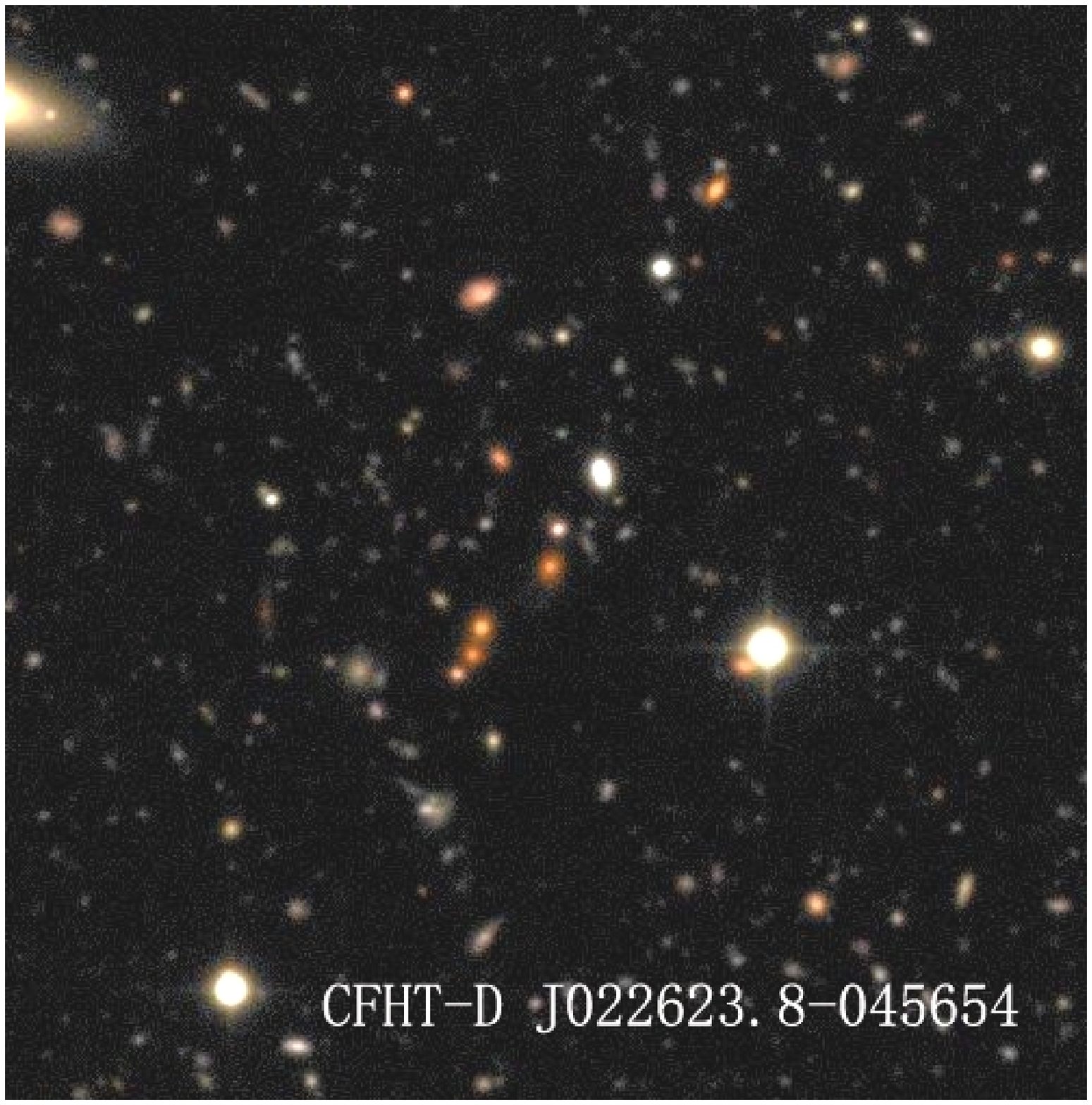}}
\resizebox{42mm}{!}{\includegraphics{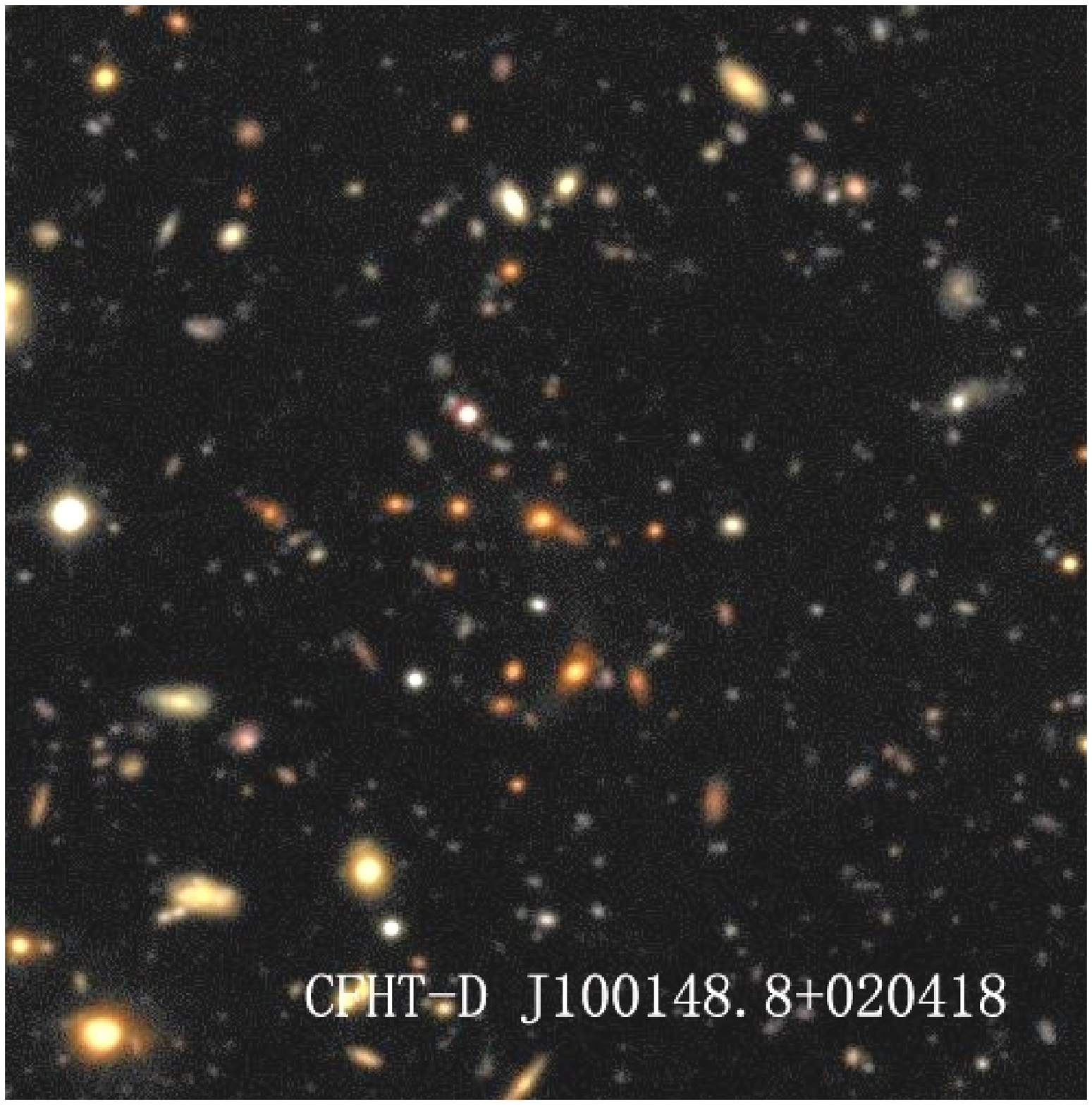}}
\caption{Examples for clusters in the CFHT Deep field identified only
  from data of \citet{iam+06} {\it (upper)} and only from data of
  \citet{cik+09} {\it (bottom)}. The images have a field of view of
  $1.5'\times1.5'$. The most of bright red galaxies in the images are
  member galaxies. A color version of this figure is available in the 
online journal.
\label{example}}
\end{figure}

As listed in Table~\ref{cat}, we get 631 clusters from the CFHT Wide
field, 187 clusters from the COSMOS field and 737 clusters from the
SWIRE field. In the CFHT Deep field, two sets of photometric redshifts
by \citet{cik+09} and \citet{iam+06} are available for galaxies with
similar uncertainties. Both are used for our cluster detection
independently. We get 163 clusters using the photo-$z$ data of
\citet{cik+09} and 105 clusters using the photo-$z$ data of
\citet{iam+06}. Among them, 66 clusters are detected from both, 97
clusters only from \citet{cik+09} and 39 only from \citet{iam+06}.
Figure~\ref{example} shows two example clusters identified only from
the data of \citet{iam+06} and two examples only from the data of
\citet{cik+09}. All clusters have an overdensity $D\ge4$ in one
dataset but less than 4 in the other. Merging two samples gives 202
unique clusters from the CFHT Deep field.

\begin{figure}
\plotone{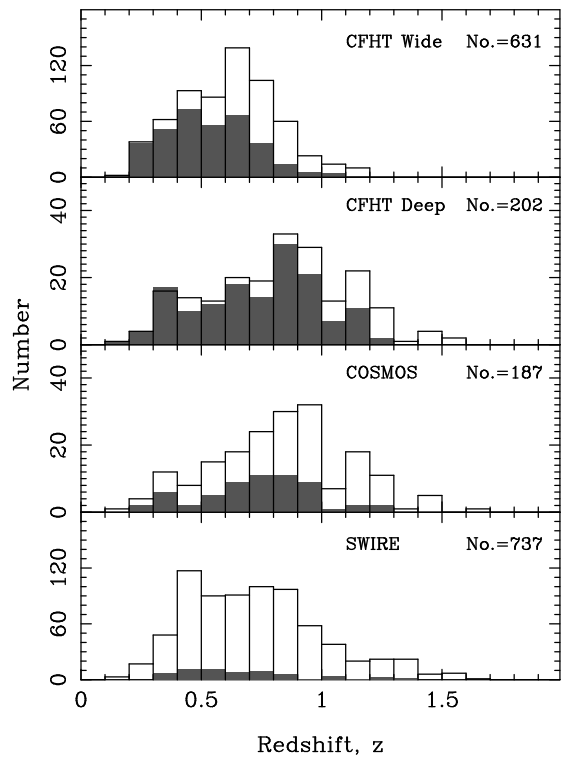}
\caption{Redshift distribution of detected clusters from the four deep
  fields. The grey area indicates the previous known clusters in our
  sample. 
\label{distz}}
\end{figure}
 
We notice that there are the overlapped regions between the four
fields: the COSMOS field is overlapped with a part (D2) of the CFHT
Deep field; a part (W1) of the CFHT Wide field is overlapped with the
D1 region of the CFHT Deep field and the XMM region of the SWIRE
field. Therefore, some clusters are repeatedly detected from the
photometric redshift data of different fields. Merging 1757 entries of
clusters in Table~\ref{cat} from the four fields gives 1644 clusters
in total. Among them, 1088 clusters are newly identified, and more
than half of them are identified from the large SWIRE field {\it which
  has not been searched for clusters previously}. Among 228 clusters
of $z\ge1$, 191 are newly identified, and again, more than half of
them are identified from the SWIRE field.  See Table~\ref{deepfields}
for the numbers of clusters identified from each field.

Figure~\ref{distz} shows the redshift distribution of detected
clusters in each field in the range of $0.1\lesssim z\lesssim
1.6$. The median
redshifts of the clusters from the CFHT Wide, the CFHT Deep, the
COSMOS and the SWIRE are 0.63, 0.86, 0.85, 0.70,
respectively. According to the redshift distribution of clusters,
clusters of $z\le0.7$ (i.e. less than the peak redshift) in the CFHT
Wide field probably have their luminous member galaxies selected with
a similar completeness. At higher redshifts, the richness is biased to
a smaller value, since only the brighter galaxies can be detected because
of the observational magnitude limit. Similarly, the redshift is
$z\le0.9$ for clusters in the CFHT Deep field, $z\le1$ for the COSMOS
field, $z\le0.8$ for the SWIRE field.

We check the companions within a radius of 2 Mpc and
$z\pm3\,\sigma_{\Delta z}$ of the detected clusters, and find that
76\% of clusters have no companion, 20\% of clusters have one
companion, and 4\% of clusters have two companions. Companions 
for a small number of clusters probably indicate superclusters
or the cluster mergers.

\subsection{False detection rate}

Because of the projection effect and the uncertainty of photo-$z$, it
is possible that a false cluster is detected by chance in the
photo-$z$ data of galaxies by the procedures above. We have to
estimate the false detection rate of our cluster detection algorithm.

\begin{figure}[bt]
\epsscale{1.0}
\plotone{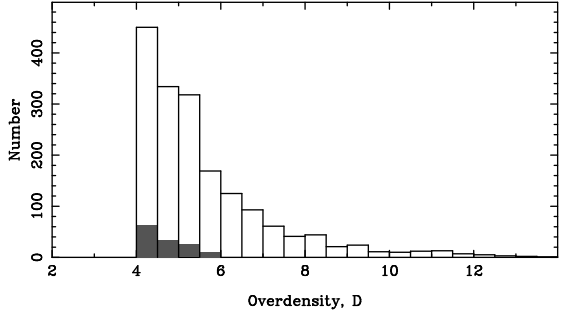}
\caption{Distribution of the overdensity of detected clusters from real 
data and the ``false'' clusters from the 3-D shuffled data (grep area).}
\label{false}
\end{figure}

Following the method of \citet{whl09}, we perform Monte Carlo
simulations based on the real data for the purpose. First, each galaxy
in the real dataset is forced to have a random walk in the sky plane
towards a random direction with a step length of a random value larger
than 1.0 Mpc but less than 2.5 Mpc. Second, we shuffle the photo-$z$
and absolute magnitude of all galaxies in each field. These procedures
should eliminate real clusters but reserve the statistical properties
of original dataset. We apply the cluster detection algorithm to such
a 3-D shuffled data, to see how many ``false clusters'' can be
detected with the criteria of $D\ge4$ and $N(0.5)\ge8$. We repeat such
tests 100 times, and get the false detection rate of 9.9\%, 2.9\%,
3.2\% and 7.9\%, respectively, for the CFHT Wide data, the CFHT Deep
data, the COSMOS data and the SWIRE data.
Figure~\ref{false} shows the average overdensity distribution of the
the detected clusters from the four fields, together with that of 
``false clusters'' from the 3-D shuffled data. The ``false'' clusters 
preferably have a low overdensity.

\subsection{The brightest cluster galaxies}
\label{ztest}

We recognize the BCG of a cluster as the brightest galaxy within a
radius of 0.5 Mpc from the cluster center and a photo-$z$ gap of
  $z\pm1.44\,\sigma_{\Delta z}$. Furthermore, we check the images of
  all clusters and BCGs to clean out any possible contamination,
  e.g., blue spiral galaxies mis-identified as BCGs and bad
  photometries. For the CFHT Wide and Deep clusters, we inspect the
  color composite CFHT images at the website\footnote{
    http://www1.cadc-ccda.hia-iha.nrc-cnrc.gc.ca/community/CFHTLS-SG
    \\ /docs/csky.html}. For the COSMOS and SWIRE clusters overlapped
  in the CFHT field, we view the CFHT images, otherwise the COSMOS
  images\footnote{ http://irsa.ipac.caltech.edu/data/COSMOS} or the
  SWIRE
  images\footnote{http://irsa.ipac.caltech.edu/data/SPITZER/SWIRE}.
  We also inspect the SDSS color composite
  images\footnote{http://cas.sdss.org/astrodr7/en} for clusters of
  $z<0.6$ located at the SDSS field. The coordinates and the AB
  magnitudes of the BCGs are given in Table~\ref{cat}. The AB magnitudes
  are taken from the data for the CFHT, the COSMOS and the XMM region
  of the SWIRE field. The Vega magnitudes are taken from that for
  the SWIRE field (except the XMM region) and are transfer to AB
  magnitude\footnote{
    http://www.sdss.org/dr7/algorithms/sdssUBVRITransform.html} for
  the rest SWIRE field via $r_{\rm AB}=r_{\rm Vega}+0.17$.

\begin{figure}
\epsscale{1.0}
\plotone{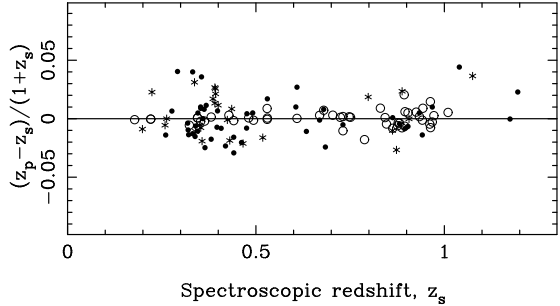}
\caption{Comparison between the cluster redshift $z_p$ (the median of
  photo-$z$ of member galaxies) and the spectroscopic redshift of 
  BCG, $z_s$: dots for 48 CFHT clusters
  with a standard deviation, $\sigma_{(z_p-z_s)/(1+z_s)}$, of 0.017,
  open circles for 48 COSMOS clusters with a standard deviation of
  0.006, and stars for 26 SWIRE clusters with a standard deviation of
  0.018.}
\label{phzc}
\end{figure}

Because BCGs are brighter than other member galaxies, some of them
have the spectroscopic redshifts obtained already, e.g. \citet{lvg+05}
for the CFHT field, \citet{llr+07} for the COSMOS field,
\citet{rbo+08} for the SWIRE field. We also check the spectroscopic
redshifts of BCGs from the SDSS DR7 database
\citep{dr7+09}. Figure~\ref{phzc} shows the difference between the
median photometric redshift of cluster members and the spectroscopic
redshift of BCGs, $(z_p-z_s)/(1+z_s)$. The standard deviations roughly
indicate the accuracy of redshift estimate for clusters.

\begin{figure}
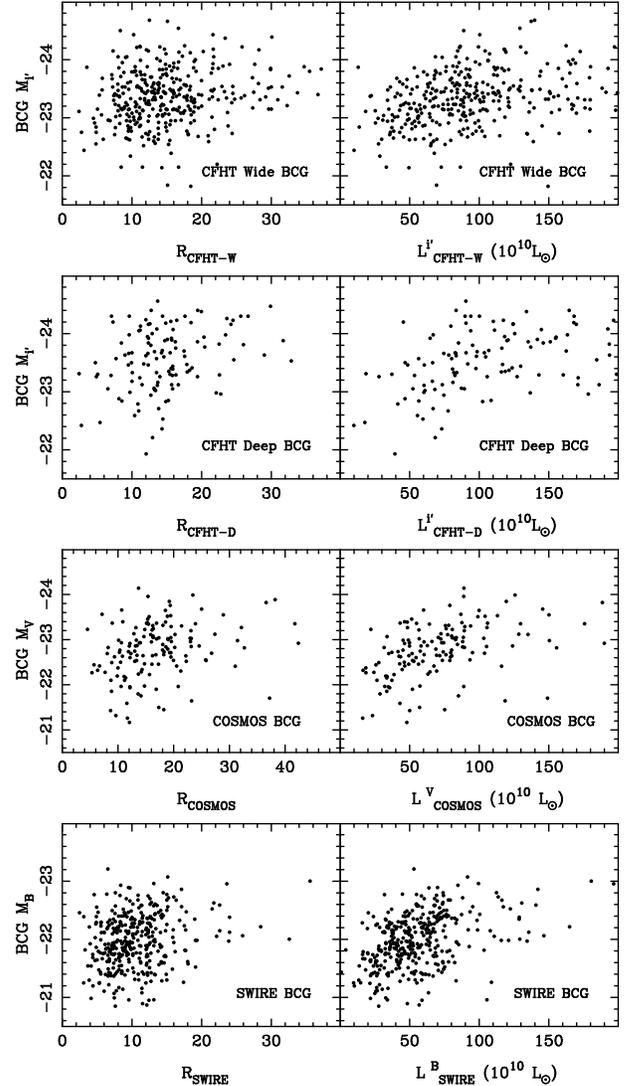

\epsscale{1.1}
\plotone{f6a.eps}\\[1mm]
\plotone{f6b.eps}\\[1mm]
\plotone{f6c.eps}\\[1mm]
\plotone{f6d.eps}
\caption{Correlations between the absolute magnitudes (i.e. the
  luminosity) of BCGs with the cluster richnesses and the total
  luminosities.}
\label{bcg_RL}
\end{figure}

The luminosities of BCGs are more or less correlated to the cluster
richness and the cluster total luminosities (see Fig.~\ref{bcg_RL}).
Richer clusters tend to have more luminous BCGs. The strongest
correlation appears between absolute magnitude of the COSMOS BCG and
total luminosity.

\subsection{Matching for different detections}
\label{scaling}

Note that the absolute magnitude limits of the member galaxy selection
are different for the four deep fields (see Table~\ref{deepfields}).
That is $M_{i'}=-21.5$ for the CFHT field, $M_V=-20.5$ of an aperture
of $3''$ for the COSMOS field, and $M_B=-20.5$ for the SWIRE
field. Different limits allow the member galaxies of different
luminosities to be selected in the cluster detection algorithm. The
richness of a cluster therefore depends on the absolute magnitude
limits.

\begin{figure}[bt]
\epsscale{1.1}
\plotone{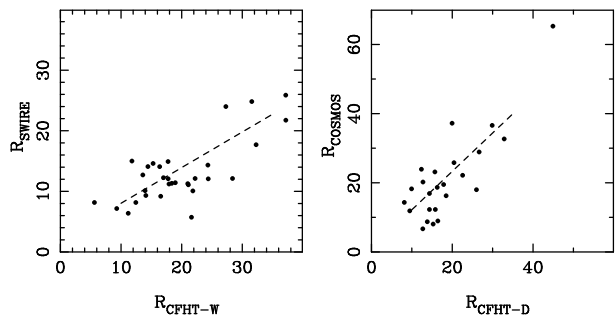}
\caption{Comparison between the richnesses of matched clusters in the
  overlapped regions between the CFHT Wide and the SWIRE fields and 
  between the CFHT Deep and the COSMOS with different absolute magnitude
  limits. 
\label{richcc}}
\end{figure}

Clusters in some overlap regions have been detected from the dataset
of two deep fields via the cluster detection algorithm, matched within
a separation of 1 Mpc and a redshift difference of $\Delta
z\le0.04(1+z)$. We compare their richnesses obtained from the two
field datasets with different absolute magnitude limits at different
bands, if luminous member galaxies are almost completely selected. We
get 6 so matched clusters in the overlapped region of the CFHT Wide
field and the CFHT Deep field, 24 matched clusters in the COSMOS and
CFHT Deep fields, 31 matched clusters in the SWIRE and CFHT Wide
fields. Figure~\ref{richcc} compares the richnesses of the matched
clusters obtained from the overlapped regions between the CFHT Wide and
the SWIRE fields and between the CFHT Deep and the COSMOS fields. As
expected, these richnesses are correlated.

\begin{figure}
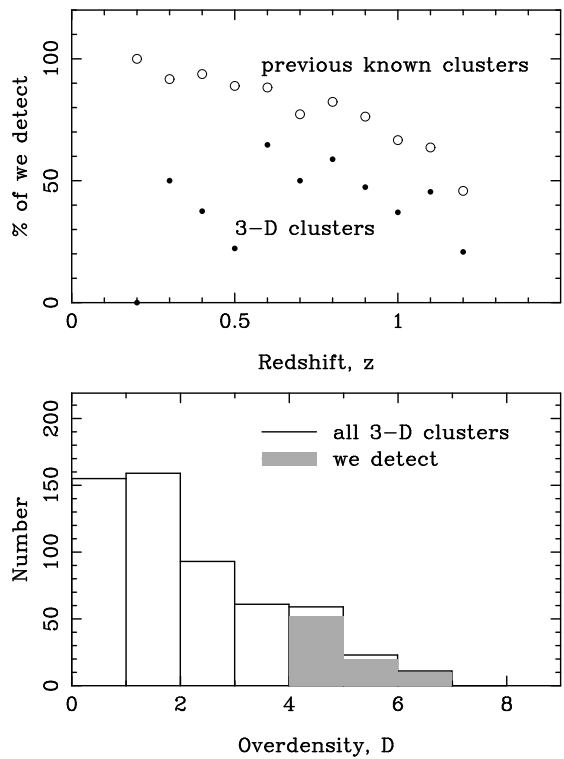

\plotone{f8a.eps}\\[2mm]
\plotone{f8b.eps}
\caption{Known clusters and our detections. {\it Upper panel:} 
  the percentage of our detected clusters are previously known in the
  CFHT Deep field (open circle), especially by 3-D methods (black
  dots) in \citet{adb+10} and \citet{mvh+10}. {\it Lower panel:} 
  the number of the 3-D clusters we detect. Apparently we detect 
  the 3-D clusters of only a high overdensity.
\label{matchf}}
\end{figure}

As shown in the Figure~\ref{distz} and mentioned in Sect.~1.1, a large
amount of clusters have previously been detected in the CFHT Wide
and Deep fields. For example,
\citet{obc+07} applied a matched-filter algorithm to the $i'$-band
four CFHT Deep fields, and identified 162 cluster candidates with an
estimated redshift uncertainty of $\sigma_{\Delta z}=0.1$.
\citet{gbm09} applied the method to the $r'$ and $z'$ bands the four
CFHT Deep fields, and found 114 and 247 cluster candidates from the
two bands, respectively. \citet[][and private
  communication]{twc09} used a red sequence method to the CFHT Wide
field of the 161 deg$^2$ data and found 5804 cluster
candidates. \citet{adb+10} used a 3-D method and identified 1029
clusters from the 28 deg$^2$ CFHT Wide field and 171 clusters from the
2.5 deg$^2$ CFHT Deep field with a mass greater than
$1.0\times10^{13}~M_{\odot}$. \citet{mvh+10} used 3D-Matched-Filter to
identify clusters from the CFHT Deep fields, and found 673 clusters in
the redshift range of $0.2\le z\le1.0$.
We merge these previous cluster samples, and get 760 unique clusters
in the CFHT Deep field, and 5848 unique clusters from the CFHT Wide
field. We match those with our sample by a separation of 1 Mpc and a
redshift difference of $\Delta z\le0.2$, and get 346 matched clusters
in the CFHT Wide field, and 144 matched clusters in the CFHT Deep
field. In Figure~\ref{matchf}, we compare the distribution of known
clusters and our detections.  At low redshifts, almost all clusters we
detect are known previously. However, we detect some new clusters
at high redshifts which are not previously known. Among the clusters
detected by the 3-D method \citep{adb+10,mvh+10}, we detect only
those rich ones in general which have a high overdensity of $D\ge4$.

\begin{figure}[bt]
\epsscale{0.8}
\plotone{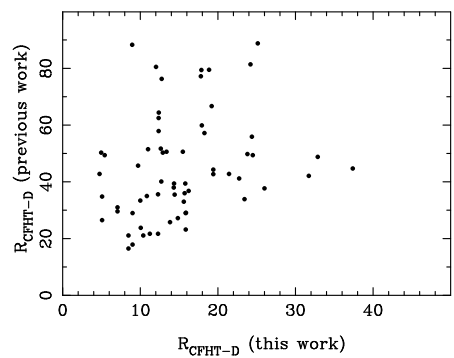}
\caption{Comparison of the richnesses of matched clusters we define
  by the number of luminous galaxies with those given in
  \citet{obc+07} and \citet{gbm09} defined as the average number of
  $L^{\ast}$ galaxies based on a single band photometric data. Here
  $L^{\ast}$ is the characteristic luminosity in the Schechter
  luminosity function.
\label{richcomp}}
\end{figure}

In Figure~\ref{richcomp}, we compare our richnesses of matched
clusters of $z\le0.9$ in the CFHT Deep field with the richness given
by \citet{obc+07} and \citet{gbm09} as the equivalent number of
$L^{\ast}$ galaxies, based on photometric data of a single band, here
$L^{\ast}$ is the characteristic luminosity in the Schechter
luminosity function. Though with a large scatter, richer clusters
found by previous authors show a larger richness by our definition.

\begin{figure}[bt]
\epsscale{1.05}
\plotone{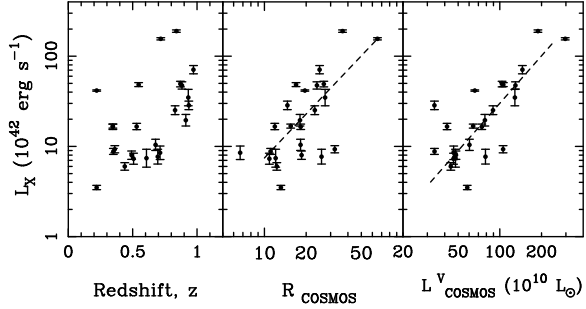}
\caption{Correlations of the X-ray luminosity with the cluster richness, 
$R_{\rm COSMOS}$, (the middle panel) and the total $V$-band luminosity, 
$L^{V}_{\rm COSMOS}$, (the right panel) with for 24 COSMOS clusters 
in the redshift range of $0.2<z<1.0$ (see the left panel).
\label{richlx}}
\end{figure}

\subsection{X-ray emission of clusters}

Some clusters in the COSMOS field have X-ray emission detected. Using
the XMM-Newton data, \citet{fgh+07} identified 72 X-ray clusters in
the 1.7 deg$^2$ of the COSMOS field, 24 of which are matched with our
sample within a projected separation of $r_p<1$ Mpc and a
redshift difference of $\Delta z\le0.05$ \citep[$\sim2\sigma$ of the
  cluster redshift uncertainty in][]{fgh+07}. These X-ray clusters
have a redshift in the range of $0.2<z<1$ (see
Figure~\ref{richlx}).

Similar to low-redshift galaxy clusters \citep{whl09}, significant
correlations are found between the X-ray luminosity of clusters and
the cluster richness ($R$, the middle panel of Figure~\ref{richlx}) or
the total $V$-band luminosity ($L_V$, the right panel of
Figure~\ref{richlx}). The best fittings to the data give
\begin{equation}
\log L^{X,42}=(-0.75\pm0.43)+(1.62\pm0.34)\log R_{\rm COSMOS},
\label{lxr}
\end{equation}
%
and
\begin{equation}
\log L^{X,42}=(-2.02\pm0.48)+(1.75\pm0.26) \log L^{V,10}_{\rm COSMOS},
\end{equation}
where $L^{X,42}$ refers to X-ray luminosity in the 0.1--2.4 keV band
in unit of $10^{42}~{\rm erg~s^{-1}}$, $L^{V,10}_{\rm COSMOS}$ refers
to the total $V$-band luminosity in unit of $10^{10}~L_{\odot}$ from
the COSMOS measurements. The correlations suggest that the richness 
we define closely relates to the properties of clusters.

\section{Stellar population and color evolution of BCGs}

The colors of BCGs provide important information on their stellar
population. Since the clusters in our sample have large redshifts, one
can easily study the stellar population and color evolution of these
BCGs. Here, we work on color evolution as a function of redshift, in
the light of a stellar population synthesis model.

\begin{figure}
\epsscale{0.93}
\plotone{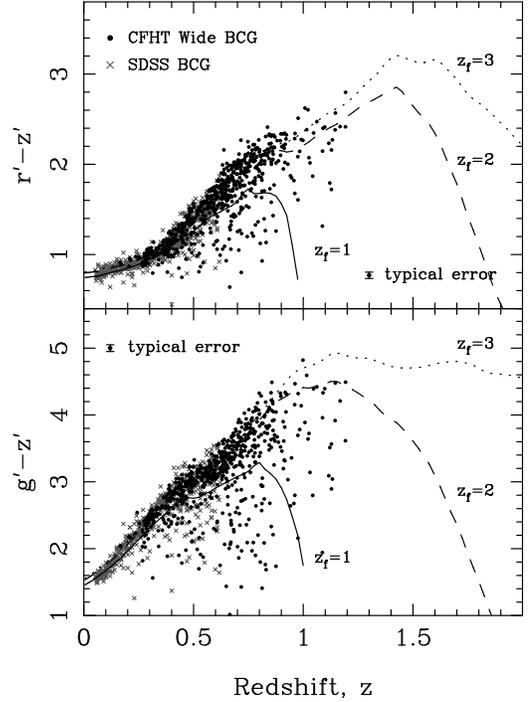}
\caption{The colors, $g'-z'$, $r'-z'$ as a function of redshift for
  BCGs in the CFHT Wide field. Three lines represent the color
  evolution with redshift from a model of passively evolved galaxies
  formed at redshift $z_f=1$ (solid line), $z_f=2$ (dashed line) and
  $z_f=3$ (dotted line), respectively. To outline the evolution at
  lower redshifts ($z<0.6$), the same-color data of 500 BCGs from the
  SDSS clusters in \citet{whl09} are included in the plots.
\label{cfhtw}}
\end{figure}

\begin{figure}[t]
\epsscale{0.93}
\plotone{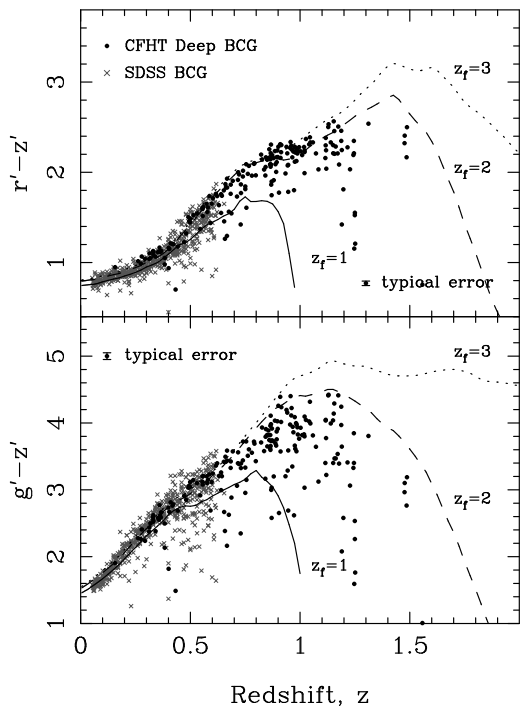}
\caption{The same as Figure~\ref{cfhtw} but for BCGs of clusters found
  in the CFHT Deep field.
\label{cfhtd}}
\end{figure}

For the CFHT data, the BCG magnitudes at $u'g'r'i'z'$ bands are
available from \citet{cik+09}. The $r'$ (7480 \AA) and $z'$ (8930 \AA)
bands correspond to the rest-frame optical wavelength for observed
BCGs up to redshift $z\sim1$, and the $g'$ band (4860 \AA) corresponds
to the rest-frame optical wavelength at low redshifts and the UV
wavelength at high redshifts. The BCG magnitudes at the $r'$, $z'$
and $g'$ bands are used to define the BCG colors for the clusters in
the CFHT Wide and Deep fields, as shown in Figure~\ref{cfhtw} and
Figure~\ref{cfhtd}. 

To explain the color changes, we show the evolution lines calculated
by using the stellar population synthesis models \citep[][BC03,
  hereafter]{bc03}. We adopt the stellar evolution tracks of Padova
1994 \citep{gbc+96}, the Basel3.1 stellar spectral library
\citep{wlb+02}, and the initial mass function of \citet{cha03} and the
Solar metallicity. The solid line in Figure~\ref{cfhtw} stands for the
color evolution for a galaxy formed at redshift $z_f=1$ and evolved
passively with cosmic time, and the dashed and dotted lines for
galaxies formed at redshift $z_f=2$ and 3, respectively.  We find that
the BCG color\footnote{When comparing the color data with model, we
  have to systematically add 0.1 to $r'-z'$ and 0.2 to $g'-z'$ of BCGs
  in the CFHT Wide field (see Figure~\ref{cfhtw}) to get consistence
  at low redshift ($z\lesssim0.3$). For the BCGs from the CFHT Deep
  field in Figure~\ref{cfhtd}, we have to systematically add 0.2 to
  $r'-z'$ and 0.3 to $g'-z'$. These calibrations are necessary to
  compensate the possible systematical bias in photometric data of
  different bands.}, $r'-z'$, is consistent with the BC03 model of
$z_f= 2$ and 3, but significant redder than that of $z_f=1$. The BCG
color, $g'-z'$, of $z\lesssim0.8$ is consistent with models of $z_f=$
2 and 3, but significant bluer than the models at higher redshift.

\begin{figure}[t]
\epsscale{0.93}
\plotone{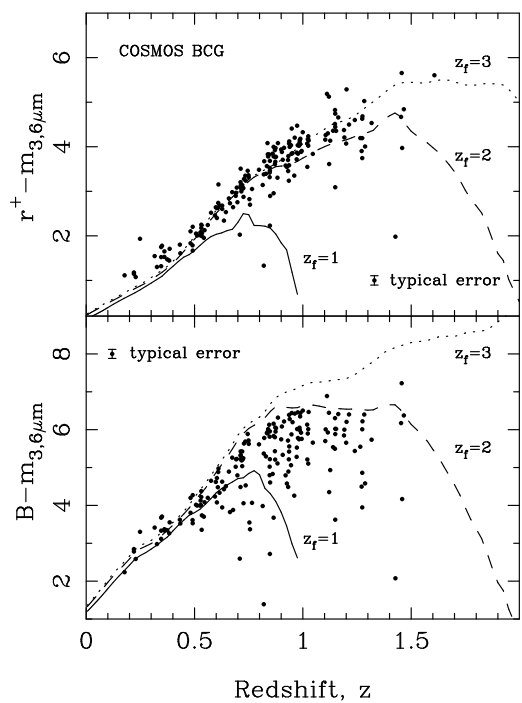}
\caption{The color evolution of BCGs in the COSMOS field.  The lines
  are calculated from the BC03 model as the same as Figure~\ref{cfhtw}
  but for the colors, $r^+-m_{3.6\mu m}$ and $B-m_{3.6\mu m}$.
\label{cosmos}}
\end{figure}

For the COSMOS data, the $B$ (4440 \AA), $r^+$ (6232 \AA) and
$m_{3.6\mu m}$ bands are used to define the BCG colors. The $B$ and
$r^+$ band data are available from \citet{ics+09}. The $B$ band
corresponds to the rest-frame optical wavelength at low redshifts, but
the rest-frame UV wavelength at high redshifts. We obtain the
$3.6\mu$m flux within an aperture of $1.9''$ diameter from the Spitzer
S-COSMOS survey \citep{ssa+07}. Since the optical fluxes are measured
over an aperture of $3''$ in diameter which encloses 75\% of the flux
for a point-like source, we therefore convert the $3.6\mu$m flux
following \citet{ics+09} to match the aperture of optical data. The AB
magnitude at $3.6\mu$m is then calculated via equation, $m_{3.6\mu
  m}=23.9-2.5\log(f_{3.6\mu m})$, where $f_{3.6\mu m}$ is in unit of
$\mu$Jy.

\begin{figure}[tbh]
\epsscale{0.93}
\plotone{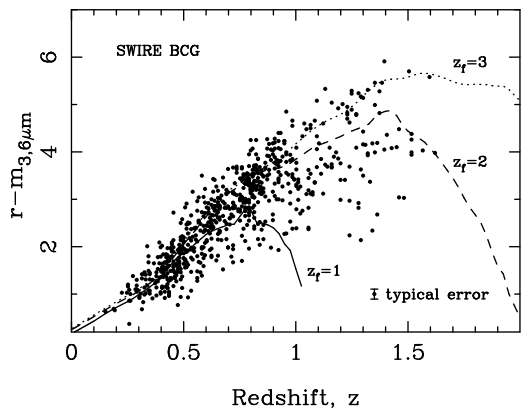}
\caption{The color evolution of BCGs in the Spitzer SWIRE field. The
  lines are calculated from the BC03 model as Figure~\ref{cfhtw} but
  for the color of $r-m_{3.6\mu m}$.
\label{swire}}
\end{figure}

Figure~\ref{cosmos} shows the color evolution of BCGs in the COSMOS
field. We find that the color, $r'-3.6\mu m$, is consistent with the
BC03 model for $z_f=$ 2 and 3. The color data of $B-m_{3.6\mu m}$ at
$z<0.5$ is consistent with that of models of $z_f=$ 2 and 3, but
significant bluer than the models at higher redshift.

For the SWIRE data, the magnitudes of $r$ and $3.6\mu m$ bands are
available from \citet{rbo+08} and are used for color evolution study
(see Figure~\ref{swire}). The result is consistent with that from the
COSMOS clusters.

The wavelenghts of $r'$, $z'$, $r^+$ and $m_{3.6\mu m}$ bands
correspond the optical or infrared band in the rest-frame, and the
observed emission of galaxies in these bands is dominated by old
stellar populations. The color evolution shown by these high redshift
data shows that the stellar population in most BCGs have been formed
at $z\ge 2$, in agreement with \citet[e.g.,][]{ses+08} and
\citet{wad+08}. The data of colors $g'-z'$ and $B-m_{3.6\mu
  m}$ at high redshifts in Figure~\ref{cfhtw}--\ref{cosmos} 
are related to the enhancement rest-frame UV color
which indicates star formation in the high redshift BCGs
\citep{dbw+10,hmd10}.

\section{Conclusions}

Using the photo-$z$ data of the four deep fields, we identified 631
clusters, 202 clusters, 187 clusters and 737 clusters, respectively,
from the CFHT Wide field, the CFHT Deep field, the COSMOS field and
the SWIRE fields. Clusters are recognized when an overdensity is $D\ge
4$ and more than eight luminous galaxies are found within a radius of
0.5 Mpc and a photo-$z$ gap. These clusters have redshifts in the
range of $0.1\lesssim z\lesssim1.6$. Because of overlapping areas
among the four fields, merging these cluster samples gives 1644
clusters, of which 1088 clusters are newly identified and 228 clusters
have a redshift $z\ge1.0$. The false detection rate is estimated to be
less than 10\%. The cluster redshift is estimated as the median
photo-z of member galaxies. Richer clusters tend to have more luminous
BCGs. The cluster richness and total luminosity are tightly related to
the X-ray luminosity.

The BCG colors and their evolution are studied for the large sample
clusters in the four fields. We compared them with a passive galaxy
evolution model using the stellar population synthesis. The color
evolution is consistent with the BC03 model in which the stars in BCGs
are formed at $z_f\ge2$ and evolved passively. The systematical
enhancement rest-frame UV color indicates star formation in these
BCGs.

\acknowledgments

We thank the referee for helpful comments, especially the 100 3-D
shuffled samples for checking the false detection rate. We are
grateful to Dr. Yanbin Yang for help of understanding BC03 model and
valuable discussions. We also thank K. Thanjavur and M. Milkeraitis
for providing their CFHT cluster samples.
The authors are supported by the National Natural Science foundation
of China (10821061 and 10833003), the National Key Basic Research
Science Foundation of China (2007CB815403).

\end{document}